\documentclass[aps,prl,twocolumn,superscriptaddress,showpacs,preprintnumbers,amssymb]{revtex4}

\usepackage{graphicx} 
\graphicspath{{ps}}

\newcommand{\mevcc}{\,{\ensuremath{\mathrm{\mbox{MeV}}/c^2}}}
\newcommand{\mevc}{\,{\ensuremath{\mathrm{\mbox{MeV}}/c}}}
\newcommand{\gevcc}{\,{\ensuremath{\mathrm{\mbox{GeV}}/c^2}}}

\newcommand{\gev}{\,{\ensuremath{\mathrm{\mbox{GeV}}}}}
\newcommand{\ee}{\ensuremath{e^+ e^-}}
\newcommand{\pp}{\ensuremath{\pi^+ \pi^-}}
\newcommand{\mm}{\ensuremath{\mu^+ \mu^-}}
\newcommand{\RM}{\ensuremath{M_{\mathrm{rec}}}}
\newcommand{\jp}{\ensuremath{J/\psi}}
\newcommand{\jpg}{\ensuremath{J/\psi \, \gamma}}
\newcommand{\co}{\ensuremath{\chi_{c1}}}
\newcommand{\ct}{\ensuremath{\chi_{c2}}}
\newcommand{\Uo}{\ensuremath{\Upsilon(1S)}}
\newcommand{\ks}{\ensuremath{K^0_S}}
\newcommand{\Ut}{\ensuremath{\Upsilon(2S)}}
\newcommand{\ccr}{\ensuremath{(c\bar{c})_{\mathrm{res}}}}

\newcommand{\etal}{{\it {et al.}}}
\newcommand{\ifb}{\ensuremath{\,\mathrm{fb}^{-1}}}
\newcommand{\res}{\ensuremath{(4.7^{+2.4}_{-1.8}{}^{+0.4}_{-0.5}) \cdot 10^{-5}}}
\newcommand{\resa}{\ensuremath{(4.7^{+2.4}_{-1.8}\text{(stat)}{}^{+0.4}_{-0.5}\text{(sys)}) \cdot 10^{-5}}}
\newcommand{\resns}{\ensuremath{(4.7^{+2.4}_{-1.8}) \cdot 10^{-5}}}

\newcommand{\gisr}{\ensuremath{\gamma_{\text {ISR}}}}
\newcommand{\gfsr}{\ensuremath{\gamma_{\text {FSR}}}}
\newcommand{\Br}{\ensuremath{\mathcal{B}}}

\begin{document}

\title{\quad \\[0.5cm] Observation of the radiative decays of \boldmath{$\Upsilon(1S)$} to \boldmath{$\chi_{c1}$}}

\noaffiliation
\affiliation{University of the Basque Country UPV/EHU, 48080 Bilbao}
\affiliation{Beihang University, Beijing 100191}
\affiliation{Brookhaven National Laboratory, Upton, New York 11973}
\affiliation{Budker Institute of Nuclear Physics SB RAS, Novosibirsk 630090}
\affiliation{Faculty of Mathematics and Physics, Charles University, 121 16 Prague}
\affiliation{Chonnam National University, Gwangju 61186}
\affiliation{University of Cincinnati, Cincinnati, Ohio 45221}
\affiliation{Deutsches Elektronen--Synchrotron, 22607 Hamburg}
\affiliation{Duke University, Durham, North Carolina 27708}
\affiliation{University of Florida, Gainesville, Florida 32611}
\affiliation{Key Laboratory of Nuclear Physics and Ion-beam Application (MOE) and Institute of Modern Physics, Fudan University, Shanghai 200443}
\affiliation{II. Physikalisches Institut, Georg-August-Universit\"at G\"ottingen, 37073 G\"ottingen}
\affiliation{SOKENDAI (The Graduate University for Advanced Studies), Hayama 240-0193}
\affiliation{Gyeongsang National University, Jinju 52828}
\affiliation{Department of Physics and Institute of Natural Sciences, Hanyang University, Seoul 04763}
\affiliation{University of Hawaii, Honolulu, Hawaii 96822}
\affiliation{High Energy Accelerator Research Organization (KEK), Tsukuba 305-0801}
\affiliation{J-PARC Branch, KEK Theory Center, High Energy Accelerator Research Organization (KEK), Tsukuba 305-0801}
\affiliation{Forschungszentrum J\"{u}lich, 52425 J\"{u}lich}
\affiliation{IKERBASQUE, Basque Foundation for Science, 48013 Bilbao}
\affiliation{Indian Institute of Science Education and Research Mohali, SAS Nagar, 140306}
\affiliation{Indian Institute of Technology Bhubaneswar, Satya Nagar 751007}
\affiliation{Indian Institute of Technology Guwahati, Assam 781039}
\affiliation{Indian Institute of Technology Hyderabad, Telangana 502285}
\affiliation{Indian Institute of Technology Madras, Chennai 600036}
\affiliation{Indiana University, Bloomington, Indiana 47408}
\affiliation{Institute of High Energy Physics, Chinese Academy of Sciences, Beijing 100049}
\affiliation{Institute of High Energy Physics, Vienna 1050}
\affiliation{INFN - Sezione di Napoli, 80126 Napoli}
\affiliation{INFN - Sezione di Torino, 10125 Torino}
\affiliation{Advanced Science Research Center, Japan Atomic Energy Agency, Naka 319-1195}
\affiliation{J. Stefan Institute, 1000 Ljubljana}
\affiliation{Institut f\"ur Experimentelle Teilchenphysik, Karlsruher Institut f\"ur Technologie, 76131 Karlsruhe}
\affiliation{Kennesaw State University, Kennesaw, Georgia 30144}
\affiliation{King Abdulaziz City for Science and Technology, Riyadh 11442}
\affiliation{Department of Physics, Faculty of Science, King Abdulaziz University, Jeddah 21589}
\affiliation{Korea Institute of Science and Technology Information, Daejeon 34141}
\affiliation{Korea University, Seoul 02841}
\affiliation{Kyungpook National University, Daegu 41566}
\affiliation{LAL, Univ. Paris-Sud, CNRS/IN2P3, Universit\'{e} Paris-Saclay, Orsay 91898}
\affiliation{\'Ecole Polytechnique F\'ed\'erale de Lausanne (EPFL), Lausanne 1015}
\affiliation{P.N. Lebedev Physical Institute of the Russian Academy of Sciences, Moscow 119991}
\affiliation{Faculty of Mathematics and Physics, University of Ljubljana, 1000 Ljubljana}
\affiliation{Ludwig Maximilians University, 80539 Munich}
\affiliation{Luther College, Decorah, Iowa 52101}
\affiliation{University of Maribor, 2000 Maribor}
\affiliation{Max-Planck-Institut f\"ur Physik, 80805 M\"unchen}
\affiliation{School of Physics, University of Melbourne, Victoria 3010}
\affiliation{University of Mississippi, University, Mississippi 38677}
\affiliation{University of Miyazaki, Miyazaki 889-2192}
\affiliation{Moscow Physical Engineering Institute, Moscow 115409}
\affiliation{Moscow Institute of Physics and Technology, Moscow Region 141700}
\affiliation{Graduate School of Science, Nagoya University, Nagoya 464-8602}
\affiliation{Kobayashi-Maskawa Institute, Nagoya University, Nagoya 464-8602}
\affiliation{Universit\`{a} di Napoli Federico II, 80055 Napoli}
\affiliation{Nara Women's University, Nara 630-8506}
\affiliation{National Central University, Chung-li 32054}
\affiliation{National United University, Miao Li 36003}
\affiliation{Department of Physics, National Taiwan University, Taipei 10617}
\affiliation{H. Niewodniczanski Institute of Nuclear Physics, Krakow 31-342}
\affiliation{Nippon Dental University, Niigata 951-8580}
\affiliation{Niigata University, Niigata 950-2181}
\affiliation{Novosibirsk State University, Novosibirsk 630090}
\affiliation{Osaka City University, Osaka 558-8585}
\affiliation{Pacific Northwest National Laboratory, Richland, Washington 99352}
\affiliation{Panjab University, Chandigarh 160014}
\affiliation{Peking University, Beijing 100871}
\affiliation{University of Pittsburgh, Pittsburgh, Pennsylvania 15260}
\affiliation{Research Center for Nuclear Physics, Osaka University, Osaka 567-0047}
\affiliation{Theoretical Research Division, Nishina Center, RIKEN, Saitama 351-0198}
\affiliation{University of Science and Technology of China, Hefei 230026}
\affiliation{Seoul National University, Seoul 08826}
\affiliation{Showa Pharmaceutical University, Tokyo 194-8543}
\affiliation{Soongsil University, Seoul 06978}
\affiliation{Sungkyunkwan University, Suwon 16419}
\affiliation{Department of Physics, Faculty of Science, University of Tabuk, Tabuk 71451}
\affiliation{Tata Institute of Fundamental Research, Mumbai 400005}
\affiliation{Department of Physics, Technische Universit\"at M\"unchen, 85748 Garching}
\affiliation{Department of Physics, Tohoku University, Sendai 980-8578}
\affiliation{Earthquake Research Institute, University of Tokyo, Tokyo 113-0032}
\affiliation{Department of Physics, University of Tokyo, Tokyo 113-0033}
\affiliation{Tokyo Institute of Technology, Tokyo 152-8550}
\affiliation{Tokyo Metropolitan University, Tokyo 192-0397}
\affiliation{Virginia Polytechnic Institute and State University, Blacksburg, Virginia 24061}
\affiliation{Wayne State University, Detroit, Michigan 48202}
\affiliation{Yamagata University, Yamagata 990-8560}
\affiliation{Yonsei University, Seoul 03722}
\author{P.~Katrenko}\affiliation{Moscow Institute of Physics and Technology, Moscow Region 141700}\affiliation{P.N. Lebedev Physical Institute of the Russian Academy of Sciences, Moscow 119991} 
  \author{I.~Adachi}\affiliation{High Energy Accelerator Research Organization (KEK), Tsukuba 305-0801}\affiliation{SOKENDAI (The Graduate University for Advanced Studies), Hayama 240-0193} 
  \author{H.~Aihara}\affiliation{Department of Physics, University of Tokyo, Tokyo 113-0033} 
  \author{S.~Al~Said}\affiliation{Department of Physics, Faculty of Science, University of Tabuk, Tabuk 71451}\affiliation{Department of Physics, Faculty of Science, King Abdulaziz University, Jeddah 21589} 
  \author{D.~M.~Asner}\affiliation{Brookhaven National Laboratory, Upton, New York 11973} 
  \author{T.~Aushev}\affiliation{Moscow Institute of Physics and Technology, Moscow Region 141700} 
  \author{I.~Badhrees}\affiliation{Department of Physics, Faculty of Science, University of Tabuk, Tabuk 71451}\affiliation{King Abdulaziz City for Science and Technology, Riyadh 11442} 
  \author{S.~Bahinipati}\affiliation{Indian Institute of Technology Bhubaneswar, Satya Nagar 751007} 
  \author{P.~Behera}\affiliation{Indian Institute of Technology Madras, Chennai 600036} 
  \author{C.~Bele\~{n}o}\affiliation{II. Physikalisches Institut, Georg-August-Universit\"at G\"ottingen, 37073 G\"ottingen} 
% ...
\collaboration{The Belle Collaboration}

\begin{abstract}
We report the first observation of the radiative decay of the \Uo\ into a charmonium state. The significance of the observed signal of $\Uo \to \gamma \co$ is 6.3 standard deviations including systematics. The branching fraction is calculated to be $\Br(\Uo \to \gamma \co)=\resa$. We also searched for \Uo\ radiative decays into $\chi_{c0,2}$ and $\eta_c(1S,\,2S)$, and set upper limits on their branching fractions. These results are obtained from a $24.9\ifb$ data sample collected with the Belle detector at the KEKB asymmetric-energy \ee\ collider at a center-of-mass energy equal to the \Ut\ mass using \Uo\ tagging by the $\Ut \to \Uo \pp$ transitions.
\end{abstract}

\maketitle
\tighten
\renewcommand{\thefootnote}{\fnsymbol{footnote}}
\setcounter{footnote}{0}

Heavy quarkonia, the nonrelativistic bound states of two heavy quarks, can be described in terms of nonrelativistic QCD (NRQCD)~\cite{Review2}. Vector quarkonia below the threshold of open-flavor production have been studied experimentally with high precision due to their high rate production in \ee\ annihilation. They decay predominantly via three intermediate gluons into multihadron final states. Calculations of such processes are complicated by soft QCD corrections which should be taken into account. Radiative decays of vector quarkonia could proceed via replacement of one gluon with a photon, or radiation of the photon in the initial or final state. While an additional photon inevitably lowers the overall branching fraction, some exclusive radiative processes can provide a much better NRQCD testing tool thanks to more reliable calculations, particularly if quarkonia are present in both initial and final states.

Although several exclusive radiative decays of quarkonia to various excitations of light mesons have been observed~\cite{uprad}, exclusive transitions between bottomonia and charmonia have not been found yet. Branching fractions of the \Uo\ radiative decays into the lower-lying charmonium states, \ccr, are expected to be at the level of $10^{-5}$, as calculated relying on NRQCD~\cite{CrosT}. In the  previous search for the bottomonium radiative decays no signal of any  even-charge-parity charmonia was found, and the obtained upper limits (UL) were at the level of $10^{-4}$~\cite{chp}.

In this paper we present a new search for the \Uo\ radiative decays into the $\chi_{cJ}$, $\eta_c(1S,\,2S)$. Unlike the previous Belle analysis based on \Uo\ data~\cite{chp}, in the present study we use the data taken at the \Ut-resonance energy and tag \Uo\ production via the $\Ut \to \Uo \pp$ transition. Although the number of tagged \Uo\ is several times smaller than the number of the directly produced \Uo\ used in the previous analysis, the tagging procedure drastically suppresses backgrounds, especially those from the processes with initial-state  radiation (ISR) or final-state radiation (FSR), which have an event topology similar to that of the signal. Moreover, two extra pion tracks increase a trigger efficiency for low-multiplicity final states of the charmonium decay.

This analysis is based on a data sample collected at the \Ut\ energy with an integrated luminosity of $24.9 \ifb$ corresponding to $(157.3\pm 3.6)\cdot 10^6$ \Ut\ mesons. In addition, off-resonance data collected below the $\Upsilon(4S)$-resonance with an integrated luminosity of $94.6\,\mathrm{\mbox{fb}}^{-1}$ are used to study continuum background. The data are collected with the Belle detector~\cite{Belle} at the KEKB asymmetric-energy \ee\ collider~\cite{KEKB}. The detector components relevant to our study are: a tracking system comprising a silicon vertex detector (SVD) and a 50-layer central drift chamber (CDC), a particle identification (PID) system that consists of a barrel-like arrangement of time-of-flight scintillation counters (TOF) and an array of aerogel threshold Cherenkov counters (ACC), and a CsI(Tl) crystal-based electromagnetic calorimeter (ECL). All these components are located inside a superconducting solenoid coil that provides a 1.5 T magnetic field. An iron flux return located outside of the coil (KLM) is instrumented to detect $K^0_L$-mesons and to identify muons.

We perform the full reconstruction of the decay chain $\Ut \to \Uo \pp$; $\Uo \to \gamma \, \ccr$, where \ccr\ are charmonia with a positive charge parity reconstructed in the following modes: $\chi_{c1,2} \to \jp(\mm)\gamma$; $\chi_{c0} \to K^+K^-\pi^+ \pi^-$; $\eta_c(1S,\,2S)\to \ks K^{\pm} \pi^{\mp}$. Thus, the final state includes a pion pair, a hard photon, and a reconstructed charmonium.

All charged tracks except for pions from \ks\ decays are required to be  consistent with originating from the interaction point. Muon and charged kaon candidates are required to be positively identified as described in Ref.~\cite{Belle}. No identification requirement is applied for pion candidates. \ks\ candidates are reconstructed by combining $\pi^+ \pi^-$ pairs with an invariant mass within $10\mevcc$ of the nominal \ks\ mass~\cite{PDG}. We require the distance between the tracks at the \ks\ vertex to be less than $1\,\mathrm{cm}$, the transverse flight distance from the interaction point to be greater than $1\,\mathrm{mm}$ and the angle between the \ks\ momentum direction and its decay path to be smaller than $0.1\,\mathrm{rad}$. We allow up to one extra charged track not included in the list of particles in the event reconstruction to account for fake, split, or pile-up background tracks. Photons are reconstructed in the electromagnetic calorimeter as showers with energy greater than 50 MeV that are not associated with charged tracks. Presence of the hard photon ($E>3\gev$) in the event is required.

The \Uo\ is tagged by the requirement on the mass recoiling against a pion pair (recoil mass):
\begin{eqnarray}
\RM(\pp)=\sqrt{(M_{\Upsilon(2S)}-E(\pp))^2-P^2(\pp)}, \nonumber
\end{eqnarray} 
where $M_{\Upsilon(2S)}$ is the \Ut\ mass, $E(\pp)$ and $P(\pp)$ are energy and momentum of the reconstructed \pp\ combination in the center-of-mass (CM) system. The \RM\ spectrum in the \Ut\ data for events containing a hard photon ($E_\gamma>3\gev$) is shown in Fig.~\ref{rmpp}\,a). The signal is well described by the shape fixed from the Monte Carlo (MC) simulation; the position of the peak is a free parameter in the fit. A small shift of the data peak with respect to the \Uo\ nominal mass~\cite{PDG}, $(0.05 \pm 0.03) \mevcc$, where the uncertainty is statistical only, is within the world average uncertainty of the $\Upsilon(1S)$ mass~\cite{PDG}. The \RM(\pp) signal window is defined as $\left| \RM(\pp) - M_{\Upsilon(1S)} \right| \,<\,10 \mevcc$. The efficiency  of this requirement is equal to 96\% according to the MC simulation.
\begin{figure}[htb] 
\includegraphics[width=0.48\textwidth]{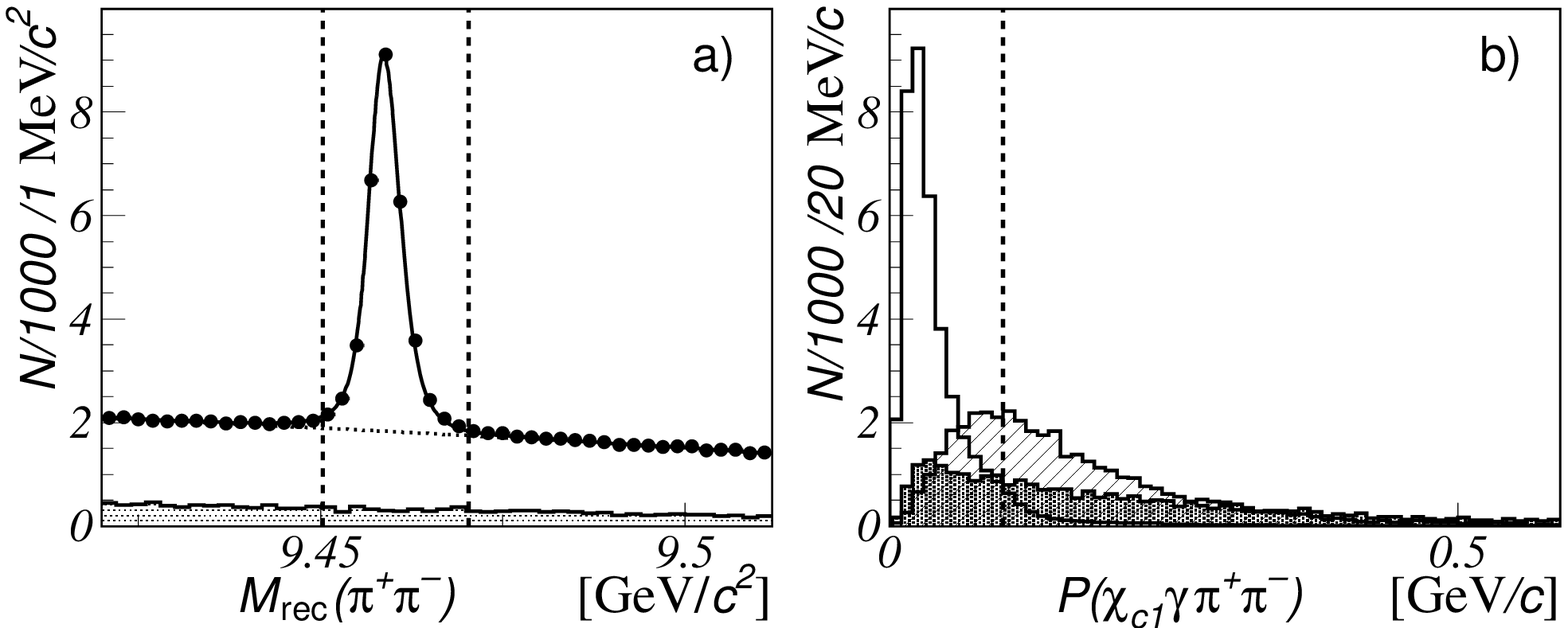} 
\caption{a) The \RM(\pp) spectrum for the data collected at the \Ut\ energy (points with errors) and expected background from $\ee \to \psi(2S) \gisr$ (histogram; not in scale). The curve is the result of the fit, with the signal shape fixed to the MC simulation, the dotted line is the background contribution. b) The distribution of the CM reconstructed momentum of \Ut\ candidates for the MC  simulated events after a mass-constrained fit (open histogram); backgrounds from the radiative return to $\psi(2S)$ and FSR $\Uo \to \mm \gfsr$ (shaded and hatched histograms). The imposed requirements are shown with the vertical dashed lines.}
\label{rmpp}
\end{figure} 

The combination of a fully reconstructed charmonium candidate and a hard photon is considered as an \Uo\ candidate. The \Uo\ mass resolution is dominated by the hard photon energy resolution, which is strongly asymmetric. The signal window is defined as $-1\,\gevcc \, < \, M(\gamma\ccr) - M_{\Upsilon(1S)} \, < \, 0.1\,\gevcc$, which covers 93\% of the signal distribution. In order to improve the momentum resolution, a mass-vertex-constrained fit of the \Uo\ candidate is performed. The \Uo\ candidate is then combined with the selected pion pair.

As all physical processes with a set of particles in the final state identical to those for the signal have a very small cross section, combinations with fake or misidentified soft charged tracks and photons are potential sources of background. In order to suppress such events, a requirement on the CM momentum of the reconstructed combination $\gamma \ccr \pp$ is applied: $P(\gamma \ccr \pp) < 100\mevc$. As demonstrated by Fig.~\ref{rmpp}\,b), the signal efficiency of this requirement is high, while the known ISR and FSR backgrounds are suppressed significantly.

We first study the decay $\Uo \to \gamma \chi_{c1,2}$; $\chi_{c1,2} \to \jpg$,  applying the criteria listed above. The \jp\ candidates are reconstructed in the dimuon mode only. The dielectron mode is not used because it is heavily  contaminated by QED processes like $\ee\to\ee\ee$ and suffers from a much lower trigger efficiency since its signature is very similar to those of radiative Bhabha events, which are intentionally suppressed by trigger requirements. The \jp\ signal region is defined as $\left| M(\mm) - M_{J/\psi} \right| \! < \!30 \mevcc$ ($\approx 2.5 \sigma$), and the sideband by the interval $[60,660]\mevcc$. The \jp\ candidates in the signal window are subjected to a mass-vertex-constrained fit, while combinations from sidebands are refitted to the center of 20 small intervals of the same width as the signal window. A $\psi(2S)$ veto is additionally imposed ($\left| M_{J/\psi \pi^+ \pi^-} - M_{\psi(2S)} \right| \, > \, 20\,\mevcc$), since the ISR process $\ee \to \psi(2S) \gisr$; $\psi(2S) \to \jp \pp$ has a large cross section and similar topology. 

The \jpg\ mass spectrum in the signal region is shown in Fig.~\ref{m_chi}\,a). Five events consistent with the $\chi_{c1}$ hypothesis are observed without any combinatorial background.

\begin{figure}[htb]
\centering \includegraphics[width=0.48\textwidth]{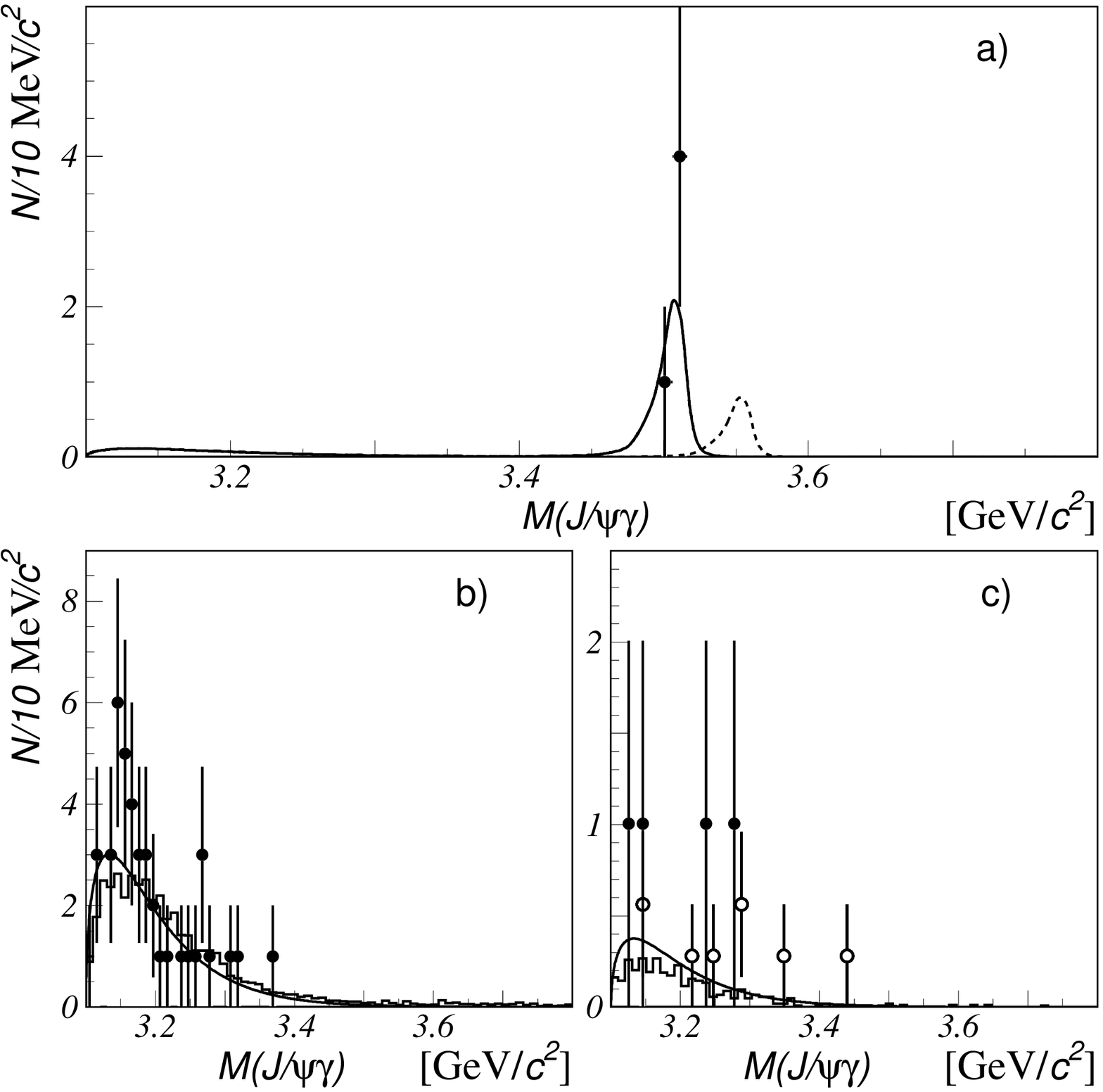}
\caption{The \jpg\ invariant mass spectrum in the \Uo\ data (closed circles with error bars): a) signal window, b) 20 times wider \jp\ mass sidebands, c) 20 times wider \RM\ sidebands; continuum data is scaled according to the ratio of luminosities and shown with open circles. Histograms are the background expectation from the MC simulation from: b) $\Uo \to \mu^+ \mu^- \gfsr$, c) $\ee \to \psi(2S) \gisr$. The solid lines show the result of the simultaneous fit to all these distributions. The dotted line shows the $\chi_{c2}$ contribution with its yield set to the 90\% confidence level UL.}
\label{m_chi} 
\end{figure}

In order to calculate the significance of the observed signal, the combinatorial background is estimated in the following categories:
\begin{enumerate}
\item [{\bf (a)}] continuum background, {\it i.e.} events other than $\ee \to \Ut$;
\item [{\bf (b)}] decays of the \Ut\ not associated with \Uo\ production;
\item [{\bf (c)}] combinatorial $\mm \gamma$ background from $\Ut \to \Uo \pp$ events;
\end{enumerate}
Non-\Uo\ backgrounds (a,~b) are studied using 20 times wider \RM\ 
sidebands: $20\,\mevcc < \left| \RM-M_{\Upsilon(1S)} \right| < 220 \,\mevcc$. The background (a) can be studied even more accurately using the continuum data sample taken below the $\Upsilon(4S)$ resonance with a 3.8 times higher integrated luminosity compared to the \Ut\ sample. The \RM\ window of 20 times larger width is used in this case: $\left| \RM - (M_{\Upsilon(1S)} - M_{\Upsilon(2S)} + \sqrt{s} ) \right| < 200\,\mevcc$. In both cases the wide \RM\ region is divided into 20 intervals of the same width as the signal one, and the \Uo\ candidate mass-constrained fit is performed to the center of the corresponding interval not to bias overall kinematics.

The \jpg\ mass spectrum for selected background events is shown in Fig.~\ref{m_chi}(c) (closed circles correspond to the \Ut\ data, open circles to the continuum data, normalized to the ratios of luminosities and energy-dependent cross sections). The numbers of observed events, 4 in the \Ut\ data, and 8 in the continuum data are in good agreement taking into account the scaling ratio (1:3.4). These numbers are also consistent with the MC expectation for the $\psi(2S)$ ISR production: MC predicts that despite a $\psi(2S)$ veto 1.8 (7.1) events would be found in the selected sample in the \Ut\ (continuum) data. Based on this study we conclude that background (b) is small in comparison with background (a). Moreover, backgrounds (a,~b) are non-peaking in the \co\ mass
region, but located in the lower invariant mass region.

The background (c) events originate from \Uo\ decays emitting energetic photons in the final state (FSR), which result in a final state similar to the one under study: $\Uo \to \mm \gfsr$. Extra soft photons to form a \co\ candidate in combination with $\mm \gamma$ originate from the next-to-leading order FSR, beam background, or pile-up. We use \jp\ sidebands to study the shape and normalization for this background source. As the \jp\ sideband candidates are refitted to the center of small intervals ($M_{\text{fit}}$), the plot of the distribution of  $M_{\mm\gamma}-M_{\text{fit}}+M_{J/\psi}$ should reproduce the shape of this background from the \jp\ signal window. This is shown in Fig.~\ref{m_chi}(b). The number of events in the 20 times wider \jp\ candidate invariant mass sidebands is $41$. The $\Uo \to \mm \gfsr$ MC simulation predicts 43 events and shows good agreement with the data in shape. We note that background (c) turned out to be dominant: $1.6 \pm 0.3$ events are expected in the signal distribution within the histogram range $[3.1,3.8]\gevcc$, to be compared with an expectation for the backgrounds (a,~b) at a level of 0.1 events.

Using MC simulation, we also estimate a possible peaking background with real \co\ produced from the ISR processes: $\ee \to X(4360,4660)\gisr$; $X(4360,4660)\to \psi(2S)\pi^+\pi^-$; $\psi(2S)\to\chi_{c1,2}\gamma$. The expected number of events from these sources is estimated to be negligibly small, $(0.9 \pm 0.1)\cdot 10^{-4}$. Another peaking background from $\Uo \to \co \pi^0$ decays with energetic $\pi^0$ decays whose clusters merge in the ECL to be misidentified as a single photon is ignored as this decay is forbidden by $C$-parity conservation.

In order to estimate the statistical significance of the observed signal, we perform a simultaneous unbinned likelihood fit to \jpg\ mass spectra in both signal and sidebands regions. The \co\ signal is described by the Crystal Ball function~\cite{cryball} with parameters fixed to the MC. Backgrounds are parameterized by the function $A\sqrt{M-M_{J/\psi}} e^{-B \cdot M}$, where $A,~B$ are free parameters. The relative normalizations of the background function in the signal, two sideband regions and continuum data are fixed according to the MC for ISR and FSR processes. The fit yields the number of signal events to be $5.0^{+2.5}_{-1.9}$, thus the estimated background contribution in the signal region is $<0.1$. We note that the background function found by the fit with free $A,~B$ parameters is in good agreement with the MC expectation both in shape and normalization. The statistical significance for the signal is defined as $\sqrt{-2\ln(\mathcal{L}_0/ \mathcal{L}_{\text {max}})}$, where $\mathcal{L}_0$ and $\mathcal{L}_{\text {max}}$ denote the likelihoods returned by the fit with the signal yield fixed at zero and at the fitted value, respectively. The significance of the \co\ signal is found to be $7.5\,\sigma$.

The reconstruction and selection efficiencies are obtained using the MC simulation. A possible effect of \co\ polarization is included in the systematic error. The total efficiency is equal to $\eta=19.2\%$, and $\Br (\Upsilon(1S) \to \gamma \co)$ is calculated according to the formula:
\begin{eqnarray}
\Br(\Uo \to \! \gamma \co ) \! = \! \frac{N_{\co}}{N_{\Ut} \eta \Br(\Ut)\Br(\co) \Br(\jp)}, 
\nonumber
\end{eqnarray}
to be $\resns$. We also set an UL on the branching fraction of the $\chi_{c2}$ production. We perform the same fit adding an extra Crystal Ball function to describe a possible $\chi_{c2}$ signal and obtain $N_{\chi_{c2}}<2.0$ at 90\%\ confidence level (CL). Finally, the branching fraction is calculated to be $\mathcal{B}(\Uo \to \gamma\chi_{c2}) < 3.3 \cdot 10^{-5}$ at 90\% CL.

The systematic errors for the \co\ significance are taken into account by assuming the most conservative background behaviour: we use the background function with longer and larger high mass tail and fix ratios of background functions in the signal and sidebands region to the highest values. The minimal significance is $6.3\,\sigma$. The systematic errors for the measured branching fraction are dominated by the track and photon reconstruction efficiency (6\%), muon identification (2\%), angular distribution of \co\ decays (5\%), fitting systematics (${}^{+0}_{-6}\%$), and uncertainty on the number of \Ut\ (1.4\%). We checked the most important sources of systematic errors using the process $\ee \to \psi(2S) \gisr$ as a control mode with almost identical kinematics. The total systematic error is estimated to be ${}^{~+9}_{-11}\%$.

We search for other charmonium states of even charge parity in \Uo\ radiative decays. The $\eta_c(1S,\,2S)$ signals can be revealed decaying to the $\ks K^\pm \pi^\mp$, while the $\chi_{c0}$ is searched for in the $K^+K^-\pi^+ \pi^-$ final state. The $\ks K^\pm \pi^\mp$ and $K^+K^-\pi^+ \pi^-$ mass spectra for the events selected with the same criteria as for the \co\ mode, are shown in Figs.~\ref{ppkk_mass}(a) and (b), respectively. 
   \begin{figure}[htb] 
   \includegraphics[width=0.48\textwidth]{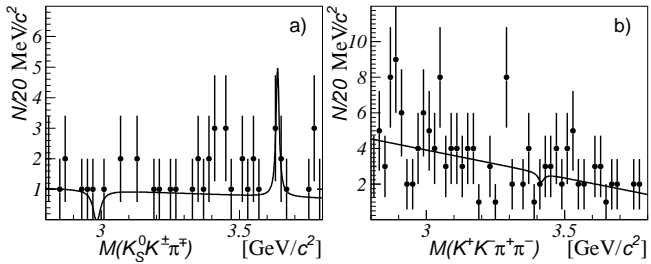}
   \caption{Invariant mass spectrum for (a) $\ks K^\pm \pi^\mp$ and (b) $K^+ K^- \pp$ modes. Histograms represent the data and curves are result of the fits described in the text.}
   \label{ppkk_mass} 
   \end{figure}    
As there are no significant peaks around expected charmonium masses (the highest significance of $\eta_c(2S)$ is $1.9 \sigma$), we set upper limits on the corresponding branching fractions. The signal functions in the fit are a Breit-Wigner function convolved with a Gaussian, with parameters fixed to the MC simulation. The backgrounds are parameterized by second-order polynomials. From the UL on the signal yields obtained by fits, we calculate the 90\% CL ULs on $\Br(\Uo \to \gamma \ccr)$ listed in Table~\ref{results}. The obtained values include systematic errors, in particular the uncertainties in the branching fractions of charmonium states into the studied modes.

\begin{table}[htb]
\caption{Summary of the measured branching fractions (in units of $10^{-5}$). The upper limits are listed at 90\% CL.}
\label{results}
\begin{center}
\begin{ruledtabular}
\begin{tabular}{l|c|c|c}
Mode & Result~~ & Previous UL~\cite{chp}~ & Prediction~\cite{CrosT}~ \\ 
\hline
\co & $4.7^{+2.4+0.4}_{-1.8-0.5}~~$ & $<2.3$ & $0.45-0.9$  \\ 
\ct & $<3.3$ & $<0.76$ & $0.51-0.56$ \\ 
$\chi_{c0}$ & $<6.6$ & $<65$ &  $0.32-0.4$ \\ 
$\eta_c(1S)$ & $<2.9$ & $<5.7$ &  $2.9-4.9$  \\ 
$\eta_c(2S)$ ~ ~ & $< 40$ & $-$  &  $-$  
\end{tabular}
\end{ruledtabular}
\end{center}
\end{table}

In summary, we report the first observation of the radiative decay of bottomonium to charmonium with $\mathcal{B}(\Uo \to \co \gamma)=\res$. We note that the obtained result is slightly higher than the previous upper limits and much higher than the theoretical expectations. However, the recent observation of \co\ production in the process $\ee \to \co \gamma$ with a large cross section~\cite{belle2} perhaps indicates a similarity of the mechanism of \co\ formation from the initial vector state with emission of photon. The new upper limits on branching fractions of other radiative decays of bottomonia to charmonia are obtained. All obtained branching fractions are summarized in the Table~\ref{results} along with the previous upper limits and the theoretical predictions.

We thank the KEKB group for excellent operation of the accelerator; the KEK cryogenics group for efficient solenoid operations; and the KEK computer group, the NII, and PNNL/EMSL for valuable computing and SINET5 network support. We acknowledge support from MEXT, JSPS and Nagoya's TLPRC (Japan);ARC (Australia); FWF (Austria); NSFC and CCEPP (China); MSMT (Czechia); CZF, DFG, EXC153, and VS (Germany); DST (India); INFN (Italy); MOE, MSIP, NRF, RSRI, FLRFAS project, GSDC of KISTI and KREONET/GLORIAD (Korea); MNiSW and NCN (Poland); RSF, Grant No. 18-12-00226 (Russia); ARRS (Slovenia); IKERBASQUE (Spain); SNSF (Switzerland); MOE and MOST (Taiwan); and DOE and NSF (USA).

\end{document}